\documentclass[a4paper]{jpconf}
\usepackage{graphicx}

\usepackage{fancybox}
\cornersize{.1}

\usepackage[numbers,square,sort&compress]{natbib}

\begin{document}
\title{Leptonic decays of $D$-wave vector quarkonia}

\author{A Krassnigg$^1$, M Gomez-Rocha$^2$ and T Hilger$^1$}

\address{$^1$Institute of Physics, University of Graz, NAWI Graz, A-8010 Graz, Austria}

\address{$^2$ECT*, Villa Tambosi, 38123 Villazzano (Trento), Italy}

\ead{andreas.krassnigg@uni-graz.at}

\begin{abstract}
We give a short and basic introduction to our covariant Dyson-Schwinger-Bethe-Salpeter-equation approach using a
rainbow-ladder truncated model of QCD, in which we investigate the leptonic decay properties
of heavy quarkonium states in the pseudoscalar and vector channels. Comparing the magnitudes of decay constants,
we identify radial $1^{--}$ excitations in our calculation with experimental
excitations of $J/\Psi$ and $\Upsilon$. Particular attention is paid to those states regarded as $D$-wave states in the quark model.
We predict $e^+ e^-$-decay width of the $\Upsilon(1^3D_1)$ and $\Upsilon(2^3D_1)$ states of the order of $\approx 15$ eV or more. 
We also provide a set of predictions for decay constants of pseudoscalar
radial excitations in heavy quarkonia. 
\end{abstract}

\section{Introduction}
The Dyson-Schwinger-Bethe-Salpeter-equation (DSBSE) approach to hadrons is a modern nonperturbative
framework based on QCD as a continuum quantum field theory \cite{Gomez-Rocha:2016cji}.
Phenomenological DSBSE studies mostly make use of the rainbow-ladder (RL) setup, since this
provides excellent value for money or, in other words, an optimal balance of richness and feasibility, e.\,g., 
\cite{Alkofer:2005ug,Eichmann:2007nn,Holl:2004yga,Eichmann:2008kk,Eichmann:2009qa,Alkofer:2009jk,Dorkin:2010pb,Blank:2010bz,Blank:2010pa,%
Mader:2011zf,Sanchis-Alepuz:2014sca,Fischer:2014xha,Fischer:2014cfa,Leitao:2015cxa,Raya:2015gva}.
Beyond RL one can explore the infinite set of Dyson-Schwinger equations (DSEs) in 
symmetry-preserving truncation schemes, see \cite{Bhagwat:2004hn,Gomez-Rocha:2014vsa,Gomez-Rocha:2015qga} and references therein. 

\section{DSBSE Setup}\label{sec:dsbse}
\begin{figure}[b]
\ovalbox{\includegraphics[height=0.118\textwidth, clip=true]{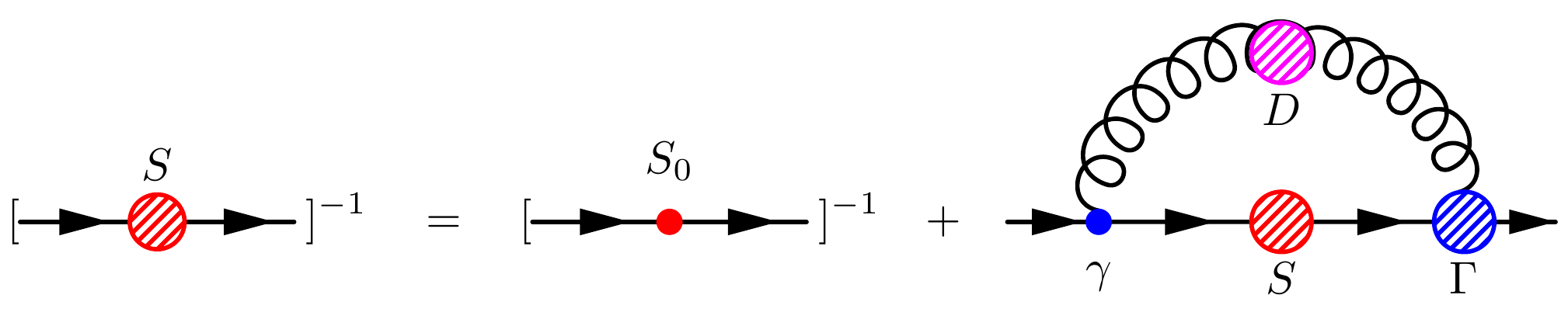}}\hfill
\ovalbox{\includegraphics[height=0.118\textwidth, clip=true]{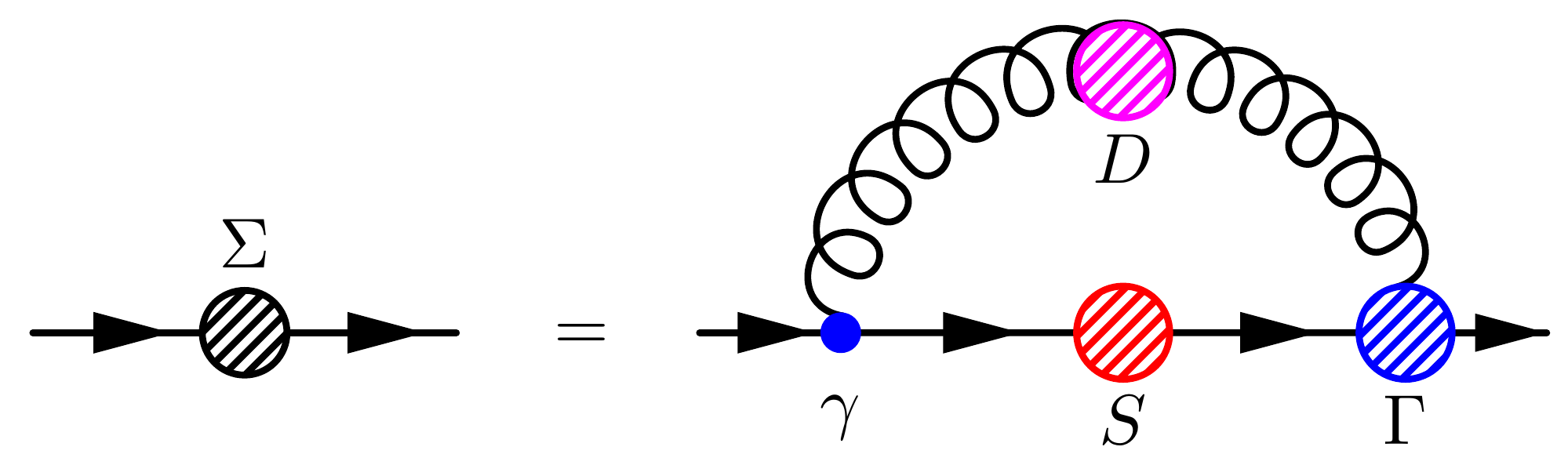}}
\caption{\label{fig:gapequation}\emph{Left panel:} quark DSE; \emph{Right panel:} quark self energy $\Sigma$.}
\end{figure}
The Dyson-Schwinger equation (DSE) for the dressed quark propagator $S$ describes how 
$S(p)$ behaves at quark momentum $p$ depending on the dressed gluon propagator $D(q)$ and the dressed quark-gluon vertex (QGV)
$\Gamma(p,q)$ (the bare QGV is denoted by $\gamma$), which is depicted in the left panel of Fig.~\ref{fig:gapequation}. 
The right panel of this figure shows the definition of the quark self energy $\Sigma(p)$ \cite{Krassnigg:2004if,Eichmann:2008ef}. 
$S_0$ denotes the free quark propagator, which serves as a
driving term in this inhomogeneous integral equation and is also used to formulate a renormalization condition for
the regularized integral \cite{Holl:2003dq}. In all figures, blobs depict dressed quantities,
while dots depict the bare counterparts. In particular, the scalar part of $S_0$ is
the current-quark mass $m_q$. Here and in the following, we assume all quantities to be renormalized
and omit indices for simplicity. Note that, due to the appearance of the inverses of $S$ and $S_0$, the quark DSE is
nonlinear.

\begin{figure}[t]
\ovalbox{\includegraphics[height=0.17\textwidth, clip=true]{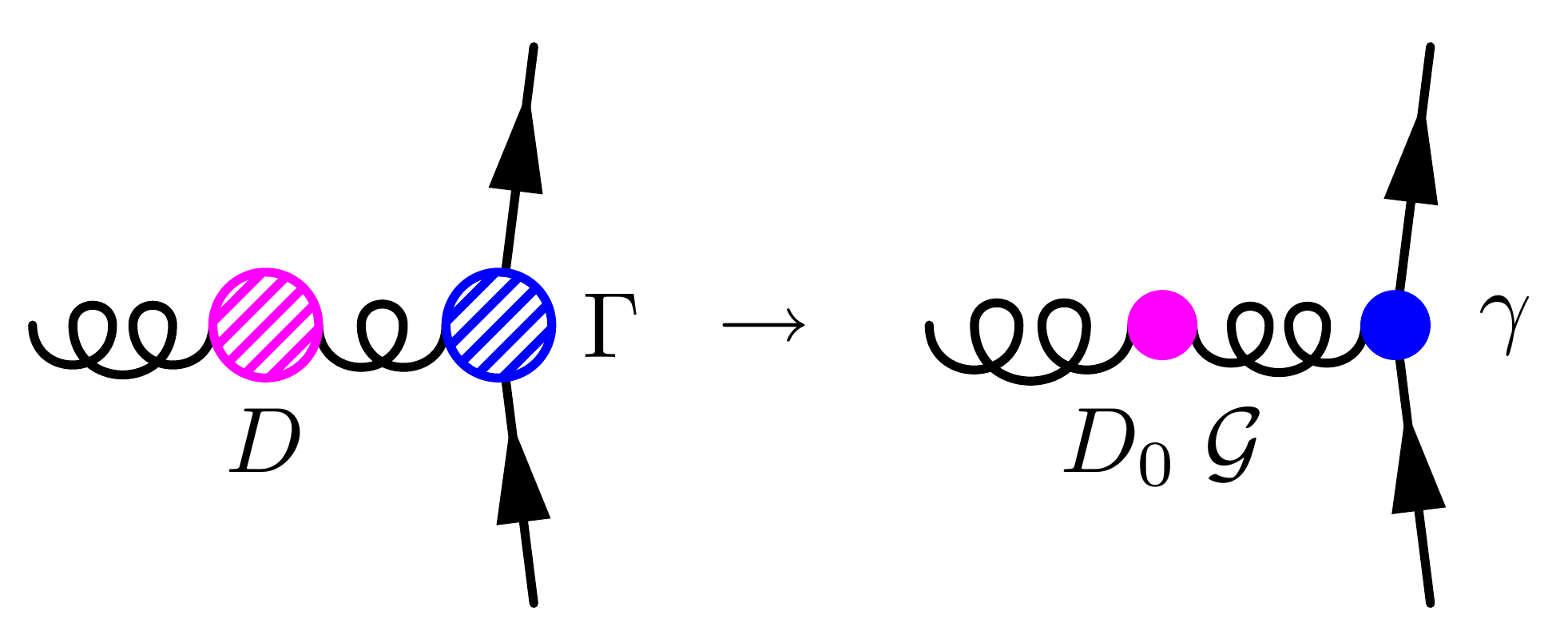}}\hfill
\ovalbox{\includegraphics[height=0.17\textwidth, clip=true]{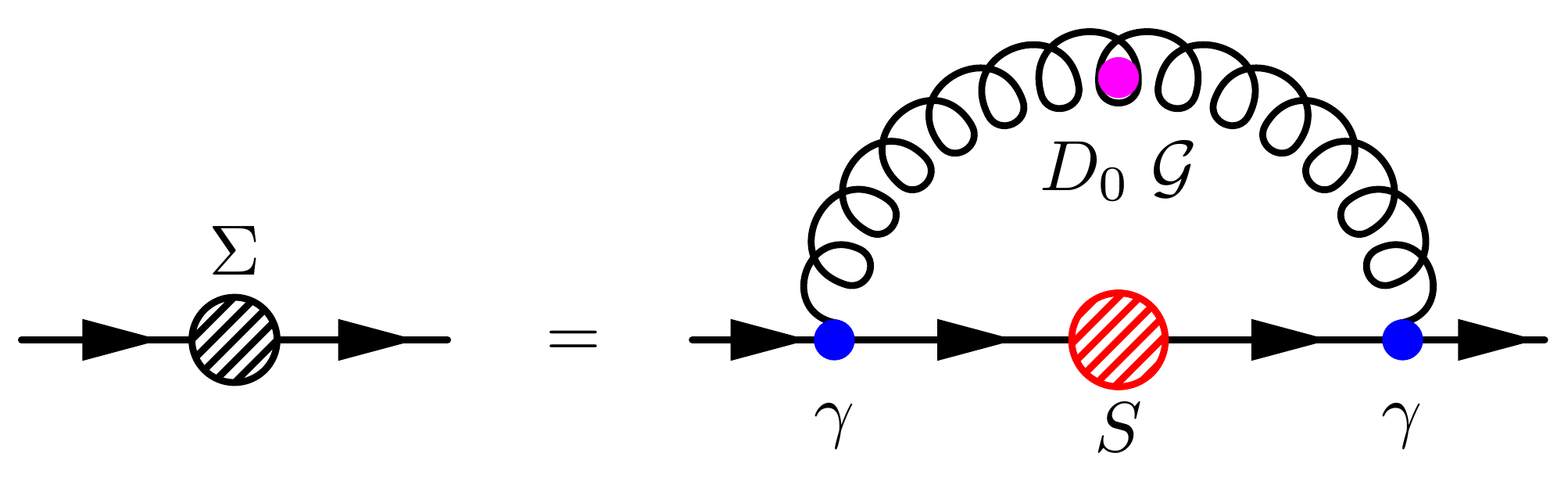}}
\caption{\label{fig:rl}\emph{Left panel:} RL \emph{Ansatz}; \emph{Right panel:} rainbow approximation of $\Sigma$ in the quark DSE.}
\end{figure}

In RL truncation one makes the \emph{Ansatz} depicted in the left panel of Fig.~\ref{fig:rl} to replace the product of
the dressed gluon propagator and dressed QGV by the product of their bare counterparts multiplied with an effective 
interaction $\mathcal{G}$ \cite{Eichmann:2008ae}. This function of the gluon momentum squared is then modeled and used to conduct
sophisticated covariant studies of hadrons. 
The right panel of Fig.~\ref{fig:rl} shows the effect of this \emph{Ansatz} in the quark self energy as defined in the right panel
of Fig.~\ref{fig:gapequation}. Basically, one is left with known quantities and can solve the equation after making
an informed choice for $\mathcal{G}$.

\begin{figure}[b]
\ovalbox{\includegraphics[height=0.16\textwidth, clip=true]{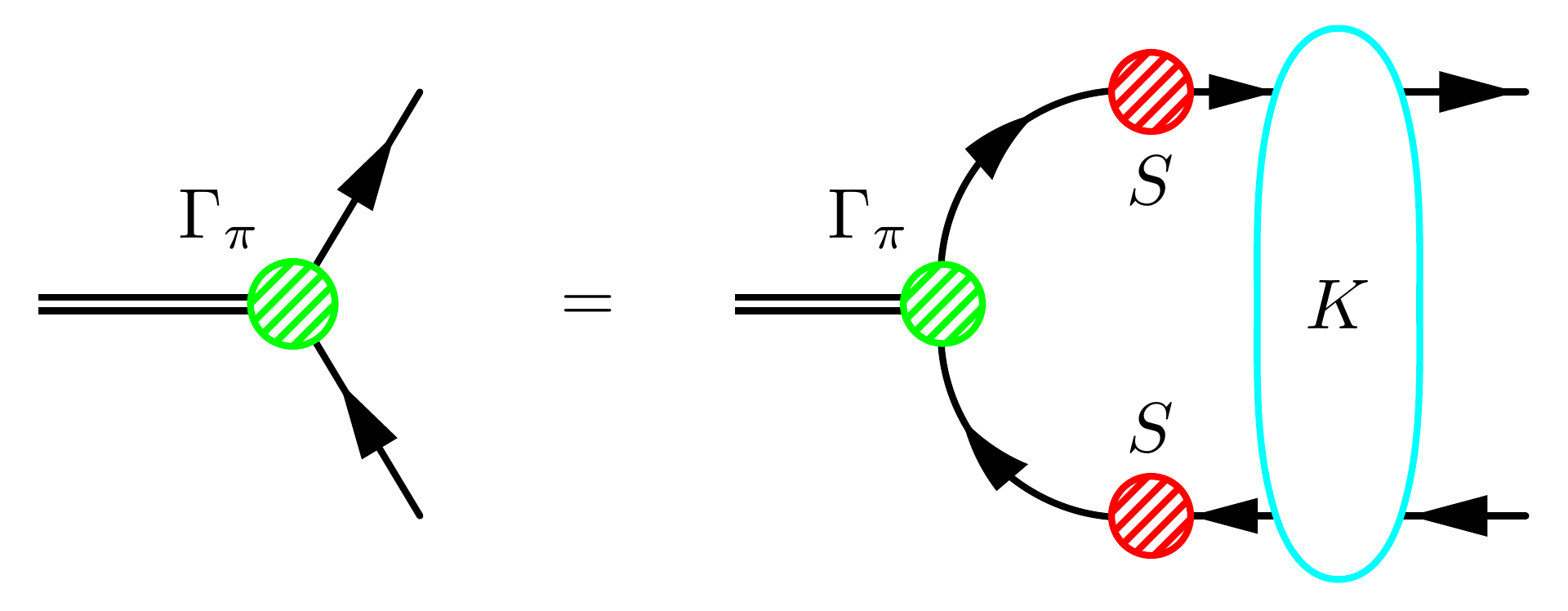}}\hfill
\ovalbox{\includegraphics[height=0.16\textwidth, clip=true]{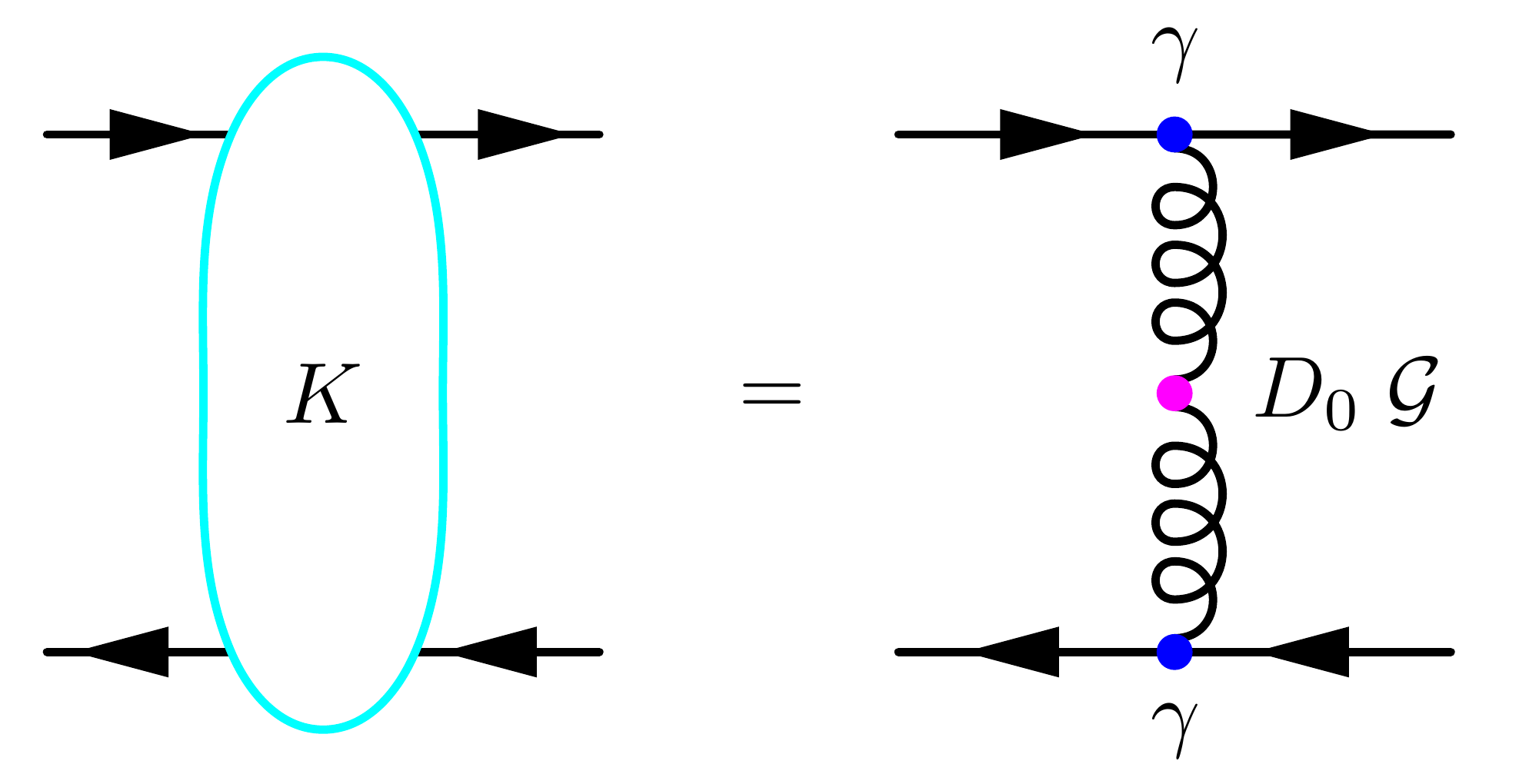}}\hfill
\ovalbox{\includegraphics[height=0.16\textwidth, clip=true]{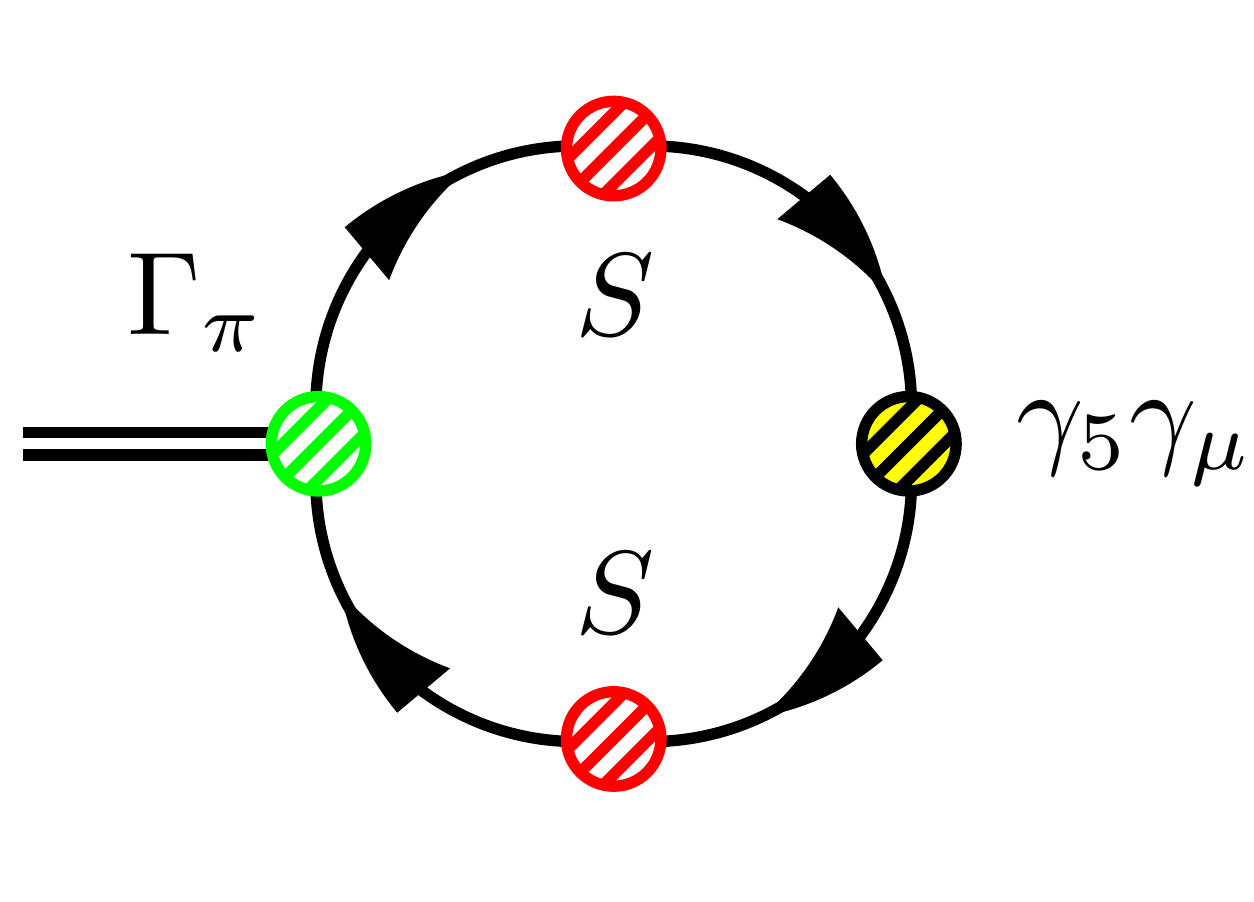}}
\caption{\label{fig:bse}\emph{Left panel:} meson BSE; \emph{Center panel:} ladder-truncated BSE kernel; \emph{Right panel:} 
diagram to compute the pseudoscalar leptonic decay constant.}
\end{figure}

The two-body bound-state equation in a quantum field theory is the Bethe-Salpeter equation (BSE), which is depicted
in the left panel of Fig.~\ref{fig:bse}. Ingredients are the dressed quark propagator, which one gets from solving
its DSE, and the amputated quark-antiquark scattering kernel $K$, which is unknown a priori \cite{Krassnigg:2009zh}. The RL truncation, inspired and
guided by the axial-vector Ward-Takahashi-identity (AVWTI), leads to an effective-gluon exchange ladder kernel as shown in 
the center panel of Fig.~\ref{fig:bse}. The AVWTI is the manifestation of chiral symmetry in the QCD context, and respecting it
guarantees a correct in-principle as well as phenomenological setup of the framework with respect to chiral symmetry and its
dynamical breaking. Concretely, RL truncation satisfies the AVWTI and thus one automatically has a massless pion in the 
chiral limit, if chiral symmetry is dynamically broken by the effective interaction \cite{Holl:2004fr,Holl:2005vu}.

The solution of the BSE is the Bethe-Salpeter amplitude (BSA) denoted in the left panel of Fig.~\ref{fig:bse} for
the pion by $\Gamma_\pi$. It is the covariant analogue of a wave function and contains all information about
the state \cite{Dorkin:2010ut,UweHilger:2012uua,Hilger:2015hka}. It is found at a particular value of one of the 
arguments of $\Gamma_\pi$, the total momentum squared $P^2$,
which gives the bound-state's mass $M$ via $M^2=-P^2$.

The ladder-truncated BSE kernel again contains only known objects such that a solution is straightforward. 
The quark DSE and meson BSE are coupled, since the DSE solution appears as input in the BSE.
We solve this coupled set of equations numerically in Euclidean momentum space. The appropriate tools are all
at hand and can be found in \cite{Bhagwat:2007rj,Krassnigg:2008gd,Blank:2010bp,Blank:2010sn,Dorkin:2013rsa,Hilger:2015zva}.
In order to calculate leptonic decay constants of mesons, one has to compute a diagram as shown in the right panel
of Fig.~\ref{fig:bse}, where the appropriate axial-vector current for $f_\pi$ is depicted as a black-yellow blob and
denoted by $\gamma_5 \gamma_\mu$ for the pseudoscalar case; the vector decay constant is obtained via $\gamma_\mu$.

\section{Results and Discussion}\label{sec:results}
\begin{figure}[t]
\includegraphics[width=0.45\textwidth, height=0.35\textwidth, keepaspectratio=false, clip=true]{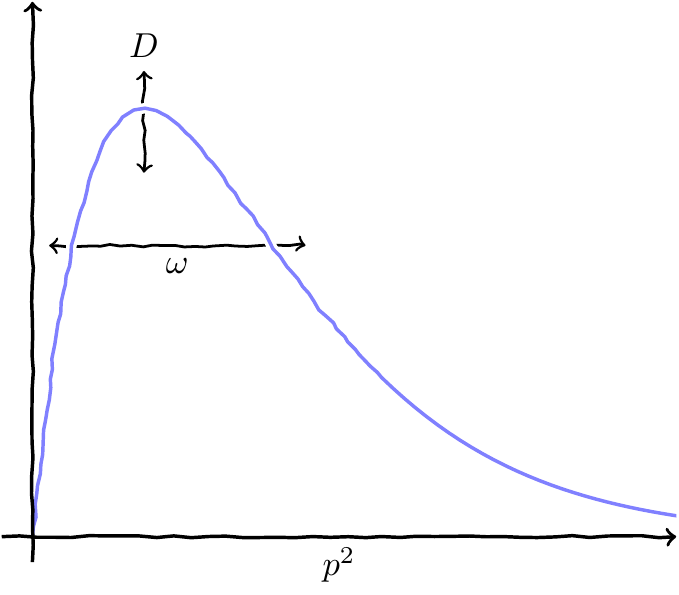}\hfill
\includegraphics[height=0.35\textwidth, clip=true]{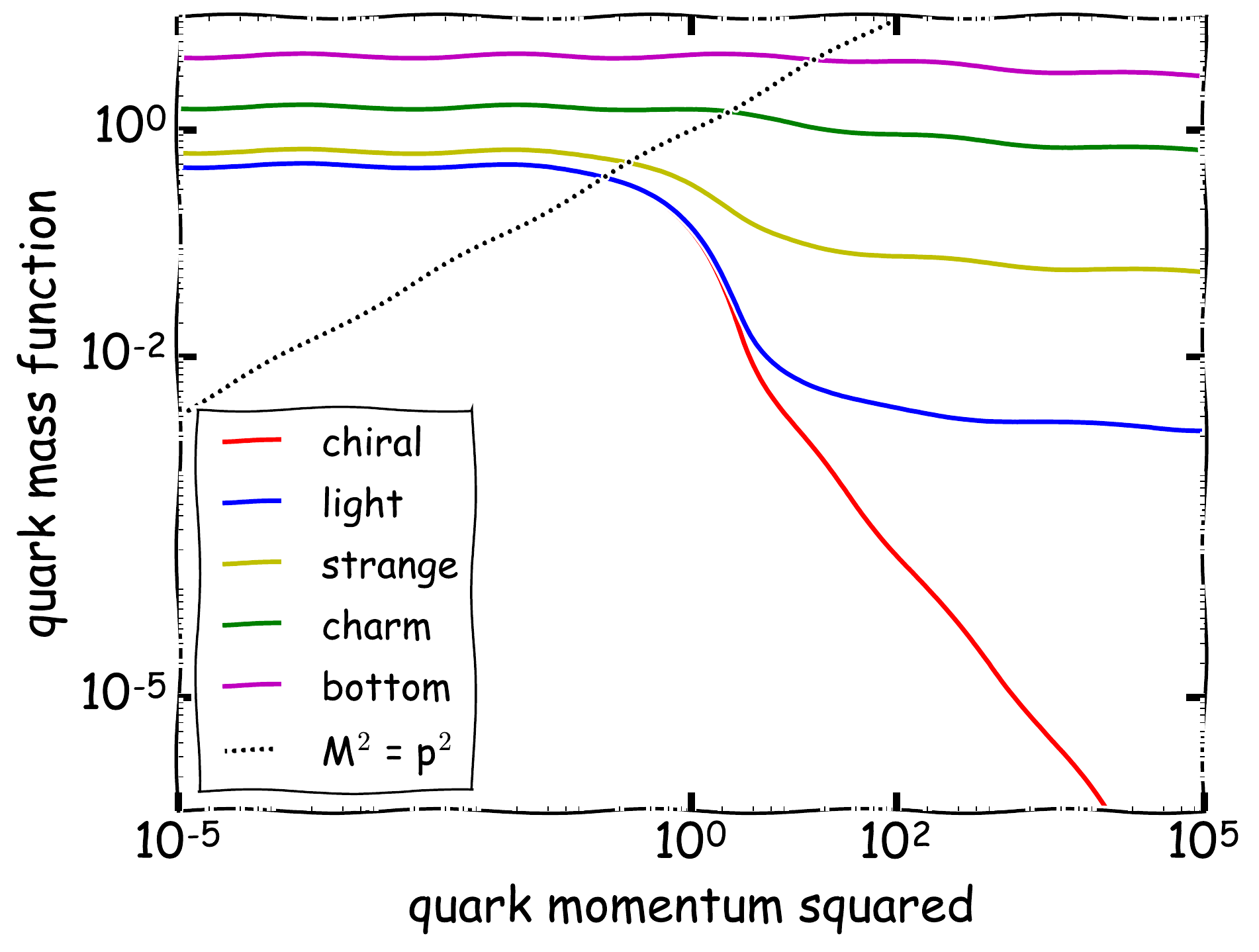}
\caption{\label{fig:dsesol}Sketches of: \emph{Left panel:} effective coupling $\mathcal{G}$; \emph{Right panel:} typical 
set of solutions of the gap equation: mass functions from the chiral limit to $b$ quark.}
\end{figure}

A very successsful and typical sophisticated effective interaction was given in \cite{Maris:1999nt}, which we use
also herein. The high-momentum behavior of the interaction is that of perturbative QCD; its shape at intermediate momenta
is determined by essentially two parameters: an inverse effective range $\omega$ and an overall strength $D$ \cite{Popovici:2014pha}. Variations
of these parameters change the interaction's shape as shown in the left panel of Fig.~\ref{fig:dsesol} \cite{Hilger:2015ora}. The right
panel of Fig.~\ref{fig:dsesol} shows a characteristic set of solutions of the quark DSE in terms of the quark mass function, where
$m_q$ assumes typical values for the masses of the up/down (light), strange, charm, and bottom quarks \cite{Hilger:2014nma}. In 
addition to those, we also show the solution in the chiral limit, i.\,e., with $m_q=0$ \cite{Hilger:2015zva}. 

The most notable feature of this set of mass functions is the transition from the perturbative domain at the right
of the plot, where one essentially deals with current quarks, to the left of the plot, where dynamical chiral symmetry breaking (D$\chi$SB)
is clearly visible in that the quarks show masses typical for corresponding constituent quarks. The border between these two
domains is $\approx 1$ GeV${}^2$, a scale set by the model parameters responsible for D$\chi$SB.

\begin{table}[t]
\caption{Calculated pseudoscalar and vector meson decay constants in MeV 
compared to experimental data \cite{Olive:2014rpp,Rosner:2005eu},
where available. \emph{Left}: charmonium; \emph{right}: bottomonium.\label{tab:dconstants}}
\begin{tabular}{llrrr} \\
State & $J^{PC}$ &            Calc. I         &    II     & Exp.  		 \\ \hline\\[-2ex]
\multicolumn{5}{l}{Pseudoscalar}  \\\hline
$\eta_c$ & $0^{-+}$ &         $401$           & $378$		    & $339(14)$  	\\
$\eta_c(2S)$ & $0^{-+}$ &     $244(12)$       & $82$		    & $189(50)$ 	\\
$\eta_c(3S)$ & $0^{-+}$ &     $145(145)$      & $206$		    & $-$ 	   		\\
$\eta_c(4S)$ & $0^{-+}$ &     $-$             & $87$		    & $-$        	\\\hline \\[-1ex]
\multicolumn{5}{l}{Vector}  \\\hline
$J/\Psi$ & $1^{--}$ &         $450$           & $411$			& $416(5)$   	\\
$\Psi(2S)$ & $1^{--}$ &       $30(3)$         & $155$			& $294(4)$   	\\
$\Psi(3770)$ & $1^{--}$ &     $118(91)$       & $45$			& $99(3)$    	\\
$\Psi(4040)$ & $1^{--}$ &     $-$             & $188$			& $187(8)$   	\\
$\Psi(4160)$ & $1^{--}$ &     $-$             & $1$			    & $142(34)$  	\\
$\Psi(4415)$ & $1^{--}$ &     $-$             & $262$			& $161(10)$  	\\\hline
\end{tabular}\hfill
\begin{tabular}{llrrr}\\
State & $J^{PC}$ &            Calc. I         &    II     & Exp. 			  \\ \hline\\[-2ex]
\multicolumn{5}{l}{Pseudoscalar}  \\\hline
$\eta_b$ & $0^{-+}$ &         $773$           & $756$			& $-$        	\\
$\eta_b(2S)$ & $0^{-+}$ &     $419(8)$        & $285$			& $-$        	\\
$\eta_b(3S)$ & $0^{-+}$ &     $534(57)$       & $333$			& $-$        	\\
$\eta_b(4S)$ & $0^{-+}$ &     $-$             & $40(15)$		& $-$        	\\\hline\\[-1ex]
\multicolumn{5}{l}{Vector}  \\\hline
$\Upsilon$ & $1^{--}$ &       $768$           & $707$			& $715(5)$   	\\ 
$\Upsilon(2S)$ & $1^{--}$ &   $467(17)$       & $393$			& $497(4)$   	\\ 
$\Upsilon(1^3D_1)$ & $1^{--}$ & $41(7)$       & $371(2)$		& $-$        	\\ 
$\Upsilon(3S)$ & $1^{--}$ &   $-$             & $9(5)$			& $430(4)$   	\\
$\Upsilon(2^3D_1)$ & $1^{--}$ &   $-$         & $165(50)$		& $-$        	\\
$\Upsilon(4S)$ & $1^{--}$ &   $-$             & $20(15)$		& $341(18)$  	\\\hline
\end{tabular}
\end{table}

After detailed studies of quarkonium spectroscopy in \cite{Blank:2011ha,Hilger:2014nma} we focus on leptonic decay constants herein.
Those were calculated for pseudoscalar and vector mesons already elsewhere, e.\,g., \cite{Maris:2006ea,Blank:2011ha,Bedolla:2015mpa},
but never systematically or 
comprehensively for radially excited states. Our results are given in Tab.~\ref{tab:dconstants} for two different sets of model
parameters and compared to experimental data where available. Set I provides the values obtained for the results as
given in \cite{Hilger:2014nma}, i.\,e., $\omega=0.7$ GeV and $D=0.5$ GeV${}^2$ for charmonium, and $\omega=0.7$ GeV and $D=1.3$ GeV${}^2$ 
for bottomonium, which were fitted separately for $\bar{c}c$ and $\bar{b}b$ to ground- and low-lying excited quarkonium states \cite{Hilger:2014nma}. 
In order to have a common parameter set for both as well as to avoid technical complications \cite{Krassnigg:2010mh} which
result in a limited number of states in set I, we present
also set II, which is uniformly obtained via  $\omega=0.3$ GeV and $D=1.3$ GeV${}^2$.

We found that our results for decay constants do not change much from one parameter set to another, so they are robust
in terms of their order of magnitude. However, some changes occur. For example, excitations can switch places as seen for the first 
and second $J/\Psi$ excitation in sets I and II. Also, the smaller the numbers get, the more sensitive they tend to be to parameter changes.
Still, on the whole one can identify the relevant patterns and we turn to set II for further analysis. By the size of the 
decay constants we match our third radial $\Upsilon$ excitation to the missing $\Upsilon(1^3D_1)$ and the fifth to the 
$\Upsilon(2^3D_1)$, thus reasonably reordering our results.

Then, our prediction for the $\Upsilon(1^3D_1)$ state, at a mass of $10.2\pm0.1$ GeV \cite{Hilger:2014nma}, is a decay
width $\Gamma(\Upsilon(1^3D_1)\rightarrow e^+ e^-)\approx 15$ eV. For the $\Upsilon(2^3D_1)$ state we obtain 
$\Gamma(\Upsilon(2^3D_1)\rightarrow e^+ e^-)\approx 75$ eV. These predictions for the decay widths are about one order 
of magnitude larger than those from other approaches, see \cite{Li:2015zda,Segovia:2016xqb} and references therein. 
A more in-depth comparison is thus necessary and will be published elsewhere.
We also note that in terms of experimental sensitivity, the upper limits for these decays are, in fact, of the order 
of $40$ eV or less \cite{Godfrey:2001vc}. 

To check the $S$/$D$-wave assignment, we investigated the contributions from various covariant structures
in the vector-meson BSA to the state's canonical norm. While $l$ is not a Lorentz invariant, we can investigate the situation
in the meson rest frame in which we solve the BSE \cite{Bhagwat:2006xi}. Upon inspection, the picture holds, i.\,e., those
states with larger leptonic decay constants receive dominant contributions to the norm from $S$-wave BSA components. In those
cases which we compare to $D$-wave states, the dominant contributions are shifted to $D$-wave BSA components with small
additional contributions from $S$-wave.
We defer a more detailed analysis and discussion of this matter to the future as well.

\ack
A.~K.~would like to thank the organizers of the FAIRNESS2016 workshop for their hospitality 
and the stimulating and productive atmosphere.
This work was supported by the Austrian Science Fund (FWF) under project no.\ P25121-N27.

\providecommand{\newblock}{}

\end{document}